\documentstyle [11pt]{article}

\begin{document}
\newtheorem{D}{Definition}[section]
\newtheorem{T}{Theorem}[section]
\newtheorem{A}{Proposition}[section]
\input epsf

\title{\bf Proof of Kolmogorovian Censorship}
\author{Gergely Bana\thanks{E-mail: gbana@hal9000.elte.hu}\\
\it{Institute for Theoretical Physics}\\
\it{E\"otv\"os University Budapest}
\\{ }
\\Thomas Durt\thanks{thomdurt@vub.ac.be}\\
\it{Department of Theoretical Physics}\\ 
\it{Vrije Universiteit Brussel}}

\maketitle
\begin{abstract}
Many argued (Accardi and Fedullo, Pitowsky) that Kolmogorov's axioms
of classical probability theory are incompatible with quantum probabilities,
and this is the reason
for the violation of  Bell's inequalities.  Szab\'o showed that, in fact, these
inequalities are not violated by
 the experimentally observed frequencies if we consider the real,
``effective'' frequencies. We prove in this work a theorem which
generalizes this result: ``effective'' frequencies associated to
quantum events always admit a Kolmogorovian representation, when these
events are collected through different experimental set ups, the
choice of which obeys a classical distribution.  
\end{abstract}
\section{Introduction}
It is commonly accepted that
the frequencies observed during the so-called Orsay experiments
(which agree with the quantum
predictions) violate the Clauser-Horne inequalities. Pitowsky (1989) 
proved that the fulfillment of these
 inequalities would be a necessary
and sufficient condition for the existence of a
Kolmogorovian representation for these frequencies (probabilities). 
Szab\'o (1995a,b) has recently shown, that the 
{\em effective 
} frequencies
observed in the Orsay experiments do no longer violate the Clauser-Horne 
inequalities. These effective frequencies are obtained
by multiplying the quantum probabilities by
the relative frequencies associated to the stochastic devices
making choice among the polarizer orientations. In an endnote of the 
article (Szab\'o
1995b), the author formulated the so-called ``Kolmogorovian Censorship" 
hypothese:
\begin{quote}
``On the basis of particular examples, it seems that there is a 
`Kolmogorovian Censorship' in the real world: We never encounter 
`naked' quantum probabilities in reality. A correlation vector consisting 
of empirically testable probabilities is always a product
\begin{eqnarray*}
 \left( p_{1}\ldots p_{n}\ldots p_{ij}\ldots \right)= \left( \pi 
_{1}\ldots 
\pi _{n}\ldots \pi _{ij}\ldots \right)\cdot \left( \tilde{p}_{1}\ldots 
\tilde{p}_{n}\ldots \tilde{p}_{ij}\ldots \right)\\= \left( \pi 
_{1}\tilde{p}_{1}\ldots \pi _{n}\tilde{p}_{n}\ldots \pi 
_{ij}\tilde{p}_{ij}\ldots \right),
\end{eqnarray*}
where  
$( \pi _{1}\ldots \pi _{n}\ldots \pi _{ij})$ 
are quantum probabilities and 
$(\tilde{p}_{1}\ldots \tilde{p}_{n}$ $\ldots  \tilde{p}_{ij})$ 
 are classical probabilities with which the corresponding measurements 
happen to be performed. My conjecture is that such a product is always 
classical. (From pure mathematical point of view, a product of a 
quantum and a classical correlation vector is not necessarily 
classical.)"
\end{quote}

 We shall first (section 3) describe the Orsay experiments and introduce 
the
formalism of Pitowsky. We shall then show how to generalize the Pitowsky 
formalism for conjunctions of more than two events. In section 4 we
prove the validity  of Szab\'o's ``Kolmogorovian 
Censorship" 
formulating it 
as a clear mathematical statement.
 \section{The Orsay experiments, the Pitowsky polytope and the 
Clauser-Horne's
inequalities.} 
\subsection{The Orsay experiments.}
The experiments realised in Orsay by Aspect et al. in order to test 
Bell's inequalities (and
also Clauser-Horne inequalities which are a variant of Bell's 
inequalities) proceed as
follows. A source emits two photons along opposite directions. Two
polarisers are placed
in two spatially separated regions (Left and Right), symmetrically, on 
both
sides of the source. A polarizer measures a dichotomic variable, the sign 
of the 
linear
polarisation of the incoming photon along a direction in the 
plane
perpendicular to its direction of propagation. The experimenter chooses, 
for
each polariser, one direction for the measurement of polarisation between 
two
different possible directions : the directions $a$ and
$a'$ in the Left region, $b$ and $b'$ in the Right
region. The technical details are not important here, but it is worth to
know  that
for some well chosen directions of the polarisers\footnote{The source 
emits a
pair of photons forming an entangled state describable by the singlet 
state, the
directions $a,\ a',\ b,\ b'$ are coplanar and are all separated by angles 
of
22.5 degrees, in the order $a',\ b',\ a,\ b$.}, we obtain by orthodox 
quantum
mechanical computations that the `naked' probabilities 
$(P(A),P(A'),P(B),P(B'),
P(A \cap B),\ P(A \cap B'),P(A' \cap B),\ P(A'\cap B'))\ $ yield $\ (1/2,
1/2,1/2,1/2,{sin^2(\pi/8)\over 2},{sin^2(\pi/8)\over 2}$,$
{sin^2(3\pi/8) \over 2},{sin^2(\pi/8)\over 2})\ $, where $A$ ($A',$ $B,$ 
$B'$)
represent the property ``the photon has + polarisation along the $a$ ($
a',$ $ b,$ $b'$) direction''
. These probabilities were
observed as experimental frequencies, with a very good precision.

In order to remain coherent with Szab\'o's notations, we shall not 
consider 
this experiment,
but a similar one where the polarisations are replaced by spins one half 
and the
polarisers are replaced by Stern-Gerlach magnets.
The initial state is the
singlet state. There are four magnets altogether (two on both sides) and 
they detect the spin-up events. 
Two 
switches, one for each particle, are making choice from sending them to 
the 
Stern-Gerlach magnets directed into different directions with 
probabilities 0.5-0.5. The observed 
events are the 
following:

\begin{center} 
\begin{tabular}{rl} 
\leavevmode
$A$ :&The ``left particle has spin `up' into direction ${\bf a} $" 
detector 
beeps\\
$A'$ :&The ``left particle has spin `up' into direction ${\bf a'} $" 
detector beeps\\
$B$ :&  The ``right particle has spin `up' into direction ${\bf b} $" 
detector beeps\\
$B'$ :& The ``right particle has spin `up' into direction ${\bf b'} $" 
detector beeps\\
$a$ :& The left switch selects direction ${\bf a} $ \\
$a'$ :& The left switch selects direction ${\bf a'} $ \\
$b$ :& The right switch selects direction ${\bf b} $ \\
$b'$ :& The right switch selects direction ${\bf b'} $ 
\end{tabular}
\end{center} 
 
For the probabilities of these events, in case of $\theta \left( {\bf 
a},{\bf a'} \right)= \theta \left( {\bf a'},{\bf b'} \right)= \theta 
\left( {\bf 
a},{\bf b'} \right)=120^{\circ } $ and $\theta \left( {\bf b},{\bf a'} 
\right)=0$, we have
\begin{eqnarray}
\label{Facts1} 
p(A)=p(A')=p(B)=p(B') &=&\frac{1}{4}  \nonumber \\
p(a)=p(a')=p(b)=p(b') &=&\frac{1}{2}  \nonumber \\
p(A\wedge a)=p(A) &=&\frac{1}{4}  \nonumber \\
p(A'\wedge a')=p(A') &=&\frac{1}{4} \nonumber \\
p(B\wedge b)=p(B) &=&\frac{1}{4}  \nonumber \\
p(B'\wedge b')=p(B') &=&\frac{1}{4}  \nonumber \\
p(A\wedge a')=p(A'\wedge a)=p(B\wedge b')=p(B'\wedge b) &=&0  
\\
p(A\wedge B)= p(A\wedge B')= p(A'\wedge B')&=&\frac{3}{32}  
\nonumber \\
p(A'\wedge B)&=&0  \nonumber \\
p(a\wedge a)= p(b\wedge b')&=&0   \nonumber \\
p(a\wedge b)= p(a\wedge b')= p(a'\wedge b)= p(a'\wedge 
b')&=&\frac{1}{4}  \nonumber \\
p(A\wedge b)= p(A\wedge b')= p(A'\wedge b)= p(A'\wedge b') 
\nonumber \\ =p(B\wedge a)= p(B\wedge a')= p(B'\wedge a)= 
p(B'\wedge a')&=&\frac{1}{8}  \nonumber 
\end{eqnarray}
These statistical data agree with quantum mechanical results, in the 
sense that
\begin{eqnarray}
\label{qprob} 
\frac{p(A\wedge a)}{p(a)}=tr(\hat{W}\hat{A})=\frac{p(A'\wedge 
a')}{p(a')}= tr(\hat{W}\hat{A'})\nonumber \\=\frac{p(B\wedge 
b)}{p(b)}=
tr(\hat{W}\hat{B})=\frac{p(B'\wedge 
b')}{p(b')}=tr(\hat{W}\hat{B'})&=&\frac{1}{2}  \nonumber \\
\frac{p(A\wedge B\wedge a\wedge b)}{p(a\wedge b)}= 
\frac{p(A\wedge B)}{p(a\wedge b)}=tr(\hat{W}\hat{A}\hat{B})  
\nonumber \\
=\frac{1}{2}\sin^{2}\frac{1}{2}\theta ({\bf a},{\bf b})&=&\frac{3}{8}  
\nonumber \\  
\frac{p(A\wedge B'\wedge a\wedge b')}{p(a\wedge b')}= 
\frac{p(A\wedge B')}{p(a\wedge b')}=tr(\hat{W}\hat{A}\hat{B'})  
\nonumber \\
=\frac{1}{2}\sin^{2}\frac{1}{2}\theta ({\bf a},{\bf b'})&=&\frac{3}{8}\\
\frac{p(A'\wedge B\wedge a'\wedge b)}{p(a'\wedge b)}= 
\frac{p(A'\wedge B)}{p(a'\wedge b)}=tr(\hat{W}\hat{A'}\hat{B}) 
\nonumber \\
=\frac{1}{2}\sin^{2}\frac{1}{2}\theta ({\bf a'},{\bf b})&=&0  \nonumber 
\\
\frac{p(A'\wedge B'\wedge a'\wedge b')}{p(a'\wedge b')}= 
\frac{p(A'\wedge B')}{p(a'\wedge b')}=tr(\hat{W}\hat{A'}\hat{B'})  
\nonumber \\
=\frac{1}{2}\sin^{2}\frac{1}{2}\theta ({\bf a'},{\bf b'})&=&\frac{3}{8}  
\nonumber 
\end{eqnarray}
where the outcomes are identified with the following projectors
\begin{eqnarray*}
\hat{A}&=&\hat{P}_{span\left\{ \psi _{+{\bf a}}\otimes \psi _{+{\bf 
a}},  \psi _{+{\bf a}}\otimes \psi _{-{\bf a}  }\right\} }\\
\hat{A'}&=&\hat{P}_{span\left\{ \psi _{+{\bf a'}}\otimes \psi _{+{\bf 
a'}},  \psi _{+{\bf a'}}\otimes \psi _{-{\bf a'}  }\right\} }\\
\hat{B}&=&\hat{P}_{span\left\{ \psi _{-{\bf b}}\otimes \psi _{+{\bf 
b}},  \psi _{+{\bf b}}\otimes \psi _{+{\bf b}  }\right\} }\\
\hat{B'}&=&\hat{P}_{span\left\{ \psi _{-{\bf b'}}\otimes \psi _{+{\bf 
b'}},  \psi _{+{\bf b'}}\otimes \psi _{+{\bf b'}  }\right\} }
\end{eqnarray*}
of the Hilbert space $H^{2}\otimes H^{2}$, and  where the singlet state 
is represented as
$\hat{W}=\hat{P}_{\Psi _{s}}$, where 
$\Psi _{s}=\frac{1}{\sqrt{2}}\left( \psi _{+{\bf a}}\otimes \psi _{-{\bf 
a} }-\psi _{-{\bf a}}\otimes \psi _{+{\bf a} }\right) $.
 
{\bf Remark} :
The probabilities
$P(A' \wedge A)$ and  $P(B \wedge B'))$ are not taken into account 
because the
choice of a direction for a Stern-Gerlach magnet excludes the other 
direction, we
cannot measure $a$ and $a'$ ($b$ and $b'$) simultaneously, the 
corresponding operators do not 
commute. We will now recall some important results of Pitowsky.

\subsection{The Pitowsky formalism} 
The question whether given probabilities are representable in 
a Kolmogorovian probability model or not can be completely 
answered. Pitowsky (1989) elaborated a convenient geometric language 
for the discussion of this problem and proved a theorem providing the 
necessary and sufficient condition of such a representation. In this 
section we recall the basic elements of his formalism and present the 
theorem. We also prove a straightforward generalization of Pitowsky's 
original 
theorem for the case of conjunctions of more than two events.

Let $S$ be a set of pairs of integers $ S\subset  \left\{ 
\left\{i,j\right\} 
\mid  1 \leq i < j \leq n\right\}$. Denote by $R(n,S)$ the linear space 
of 
real vectors having a form like ($ f_1,f_2,...,$ $f_n,...,f_{ij},...$), 
$\{i,j\}\in S
$. For 
each 
$\varepsilon \in \{0,1\}^n$, let $u^{\varepsilon }$  be the following 
vector 
in $R(n,S)$:
\begin{eqnarray*}
u^{\varepsilon}_i=\varepsilon_i, \hskip 1cm 1\leq i\leq n,\\
u^{\varepsilon}_{ij}=\varepsilon_i\varepsilon_j, \hskip 1cm \{i,j\}\in 
S.
\end{eqnarray*}

\begin{D}
\label{ D-C(n)}
The {\em classical correlation polytope } ${\cal C}(n,S)$ is the closed 
convex hull of vectors 
$\left\{u^{\varepsilon}\right\}_{\varepsilon \in 
\{0,1\}^n}$ in  $R(n,S)$: 
\begin{eqnarray*}
{\cal C}(n,S):=\\ \left\{ a\in R(n,S)\mid  a=\sum_{\varepsilon \in 
\{0,1\}^n} 
\lambda_{\varepsilon}u^{\varepsilon}, \ where\ \lambda_{\varepsilon} 
\geq 0\ 
and\ 
\sum_{\varepsilon \in\{0,1\}^n} \lambda_{\varepsilon}=1 \right\} 
\end{eqnarray*}
\end{D}
Consider now events $A_{1}, A_{2}, \ldots A_{n}$  and  some of 
their conjunctions $A_{i}\wedge A_{j}\: \left( \{i,j\}\in S\right) $. 
Assume that we can associate probabilities to them (that is, we order 
numbers to them about which we 
think that they could be probabilities), from which we can 
form a so 
called correlation vector:
\begin{eqnarray*}
& &{\bf p}=\left( p_{1}, p_{2},\ldots p_{n}, \ldots p_{ij}, \ldots 
\right) 
\\&=&\left(p(A_{1}), p(A_{2}), \ldots, p(A_{n}), \ldots p(A_{i}\wedge 
A_{j}), \ldots \right)\in R(n,S)
\end{eqnarray*}
We will then say that
\begin{D}
\label{KolmRep}
$ {\bf p} $ {\em has a Kolmogorovian 
representation } if there exist a Kolmogorovian probability space 
$(\Omega, \Sigma, \mu)$ and measurable subsets 
$$X_{A_{1}}, X_{A_{2}}, \ldots X_{A_{n}}\in \Sigma $$ 
such that 
\begin{eqnarray*}
p_{i}=\mu \left(X_{A_{i}}\right),\hskip 1cm&1\leq i\leq n,\\
p_{ij}=\mu \left(X_{A_{i}}\cap X_{A_{j}}\right),\hskip 1cm&
\{i,j\}\in S.
\end{eqnarray*}
\end{D}
The following theorem due to Pitowsky (Pitowsky 1989) allows us to 
formulate the existence of a
Kolmogorovian representation for a correlation vector in terms of a 
geometrical condition.

\begin{T}
\label{PitowskyTheorem}
A correlation vector $ {\bf p} =\left( p_{1}, p_{2},\ldots p_{n}, \ldots 
p_{ij}, \ldots \right)$ has a Kolmogorovian representation if and only if 
this vector belongs to the classical polytope (${\bf p}\in {\cal C}(n,S)$).
\end{T}

In case $n=4$  and $S=S_{4}=\left\{ \{1,3\}, \{1,4\}, \{2,3\}, 
\{2,4\}\right\} $, the condition $ {\bf p} \in {\cal C}(n,S)$ can be 
shown ( 
Pitowsky 1989) to
be equivalent to the following inequalities: 
\begin{eqnarray}
\label{Eq-Clauser_Horne} 
& &0\leq p_{ij}\leq p_{i}\leq 1,  \nonumber \\
& &0\leq p_{ij}\leq p_{j}\leq 1, \hskip 1cm i=1,2\; \;j=3,4  \nonumber 
\\
& &p_{i}+p_{j}-p_{ij}\leq 1,  \nonumber \\
& &-1\leq p_{13}+ p_{14}+ p_{24}- p_{23}- p_{1}- p_{4}\leq 0,\\
& &-1\leq p_{23}+ p_{24}+ p_{14}- p_{13}- p_{2}- p_{4}\leq 0,  
\nonumber \\
& &-1\leq p_{14}+ p_{13}+ p_{23}- p_{24}- p_{1}- p_{3}\leq 0,  
\nonumber \\
& &-1\leq p_{24}+ p_{23}+ p_{13}- p_{14}- p_{2}- p_{3}\leq 0.  
\nonumber 
\end{eqnarray}
These last four equations are equivalent to the well known 
Clauser-Horne inequalities (Clauser, Horne 1974), which are a variant of 
Bell's inequalities (Bell 1964).

\subsection{The generalized Pitowsky theorem}
\label{gpt}
Pitowsky's original theorem deals with simple conjunctions only. We 
present here a
straightforward generalization of the theorem for the  case where 
conjunctions of not only
two but three or more events are considered. 

A typical correlation vector is then: 
\begin{eqnarray}
\label{gencor} 
\left( 
p _{1},\ldots,p _{n},\ldots,p _{i_{1}i_{2}},\ldots,p _{j_{1} j_{2} 
j_{3}},\ldots p _{k_{1}k_{2}k_{3}k_{4}}\ldots \right)
\end{eqnarray}
 where  $ p _{j_{1} j_{2} j_{3}},\, p _{k_{1}k_{2}k_{3}k_{4}}$ etc. 
stand for probabilities of conjunctions of three or more events. More 
precisely, consider a set $S$ of subsets of indexes  
$ S\subset  2^{ \left\{ 1,\ldots,n\right\}}\setminus \left\{ \emptyset 
\right\} ,$ where we denote by $2^{ \left\{ 1,\ldots,n\right\}}$ the 
power 
set of $
\left\{ 1,\ldots,n\right\}.$

{\bf Remark}: In the formalism of Pitowsky, the set ($S$) of indices
related to the conjunctions does not contain the set of n pure indices (i
from 1 to n). In the notation introduced here, $S$ contains also  the set
of isolated indices i (i from 1 to n) related to one property only. This
way the notations become simpler and one does not necessarily have to 
have assumptions 
for the probabilities of $A_i$-s.

Consider the linear
space of real functions
\begin{eqnarray*}
{\bf R} (n,S)=\left\{ {\bf f}\mid  S\ni I \mapsto {\bf f}_{I}\in {\bf R} 
\right\}
\end{eqnarray*}
For each 
$\varepsilon \in \left\{ 0,1\right\}^{n}$ we define ${\bf 
u}^{\varepsilon }\in {\bf R}(n,S)  $ 
as
$${\bf u}^{\varepsilon }_{I}=\prod_{i\in I}\varepsilon _{i}, \hskip 1cm 
(\forall\: I\in S)$$ 
\begin{D}
\label{ D-C(n)}
The {\em generalized classical correlation polytope } ${\cal {\bf C} 
}(n,S)$ is the closed convex hull of vectors 
$\left\{{\bf 
u} 
^{\varepsilon}\right\}_{\varepsilon \in 
\{0,1\}^n}$ in  ${\bf R}(n,S)$: 
\begin{eqnarray*}
{\cal {\bf C} }(n,S):=\\ \left\{ {\bf a}\in {\bf R}(n,S)\mid  
{\bf a}=\sum_{\varepsilon \in 
\{0,1\}^n} 
\lambda_{\varepsilon}{\bf u}^{\varepsilon}, \   where\ 
\lambda_{\varepsilon} \geq 0\ and\ 
\sum_{\varepsilon \in\{0,1\}^n} \lambda_{\varepsilon}=1 \right\} 
\end{eqnarray*}
\end{D}
Consider now conjunctions $\bigwedge_{i\in I} A_{i}\: \left( I\in 
S\right) 
$ of events $A_{1}, A_{2}, \ldots A_{n}$. If we associate probabilities 
to them (that is numbers about which we think that they can be the 
probabilities of the events but they 
don't have to come from some well defined probability theory) we 
can form a generalized correlation vector $\left({\bf p},\{A_i 
\}_{i=1}^n\right)_S$ 
where ${\bf p}\in {\bf R}(n,S) $:
\begin{eqnarray}
\label{val}
{\bf p}_{I}=p\left( \bigwedge_{i\in I}A_{i}\right)  
\end{eqnarray}
In the following we will sometimes refer to $\left({\bf p},\{A_i 
\}_{i=1}^n\right)_S$ only by ${\bf p}$.
\begin{D}
\label{KolmRep2}

A generalized correlation vector $\left({\bf p},\{A_i 
\}_{i=1}^n\right)_S$ 
{\em has a Kolmogorovian 
representation } if there exist a Kolmogorovian probability space 
$(\Omega, \Sigma, \mu)$ and measurable subsets 
$$X_{A_{1}}, X_{A_{2}}, \ldots X_{A_{n}}\in \Sigma $$ 
such that 
\begin{eqnarray*}
{\bf p}_{I} =\mu \left( \bigcap_{i \in I} X_{A_{i}}\right),\hskip 1cm&
\ (\forall \: I\in S).
\end{eqnarray*}
We denote the representation with 
$\left(\Omega,\Sigma,\mu,\left\{X_{A_i}\right\}^n_{i=1} \right)$
\end{D}

Now we can formulate the generalization of Pitowsky's theorem and prove 
it by a straightforward 
generalisation of the original proof (Pitowsky 1989).

\begin{T}
\label{GenPit}
A generalized correlation vector ${\bf p}\in {\bf R}(n,S) $ has a 
Kolmogorovian representation if and only if it belongs to the generalised 
classical
polytope (${\bf p}
\in {\cal {\bf C}  }(n,S)$).
\end{T}
\paragraph{Proof} 
Assume,  ${\bf p}$ has a Kolmogorovian representation 
$(\Omega, \Sigma, \mu \left\{X_{A_i}\right\}_{i=1}^n)$.  
For any $\varepsilon$ in $\left\{ 0,1\right\}^{n}$
we define a measurable subset $X^{\varepsilon}=
\bigcap_{0 \leq i \leq n} X_{A_{i}}^{\varepsilon _{i}}$, where 
$X_{A_{i}}^1
= X_{A_{i}}\ ;\ X_{A_{i}}^0=\Omega \setminus X_{A_{i}}$.
One can easily check that if $\varepsilon^1,\varepsilon^2 \in \left\{ 
0,1\right\}^{n}$, $\varepsilon^{1} 
\neq \varepsilon^{2} $ 
then $X^{\varepsilon^{1} }\cap X^{\varepsilon^{2}}=\emptyset    $ , 
$\bigcup_{\varepsilon \in \left\{ 0,1\right\}^{n} }X^{\varepsilon } 
=\Omega$. Moreover, for all $I\in S$,   
\begin{eqnarray*}
\bigcap_{i\in I}X_{A_{i}} =
\bigcup_{\begin{array}{c}
\varepsilon \in \left\{ 0,1\right\}^{n}\\ \varepsilon_i =1\,\;{\rm if} 
 
\,\; 
i\in I
\end{array} }
 X^{\varepsilon}
\end{eqnarray*}
 Therefore 
\begin{eqnarray*}
{\bf p}_{I}= \mu(\bigcap_{i\in I}X_{A_{i}}) =
\sum_{\begin{array}{c}
\varepsilon \in \left\{ 0,1\right\}^{n} \\
\varepsilon_i =1\,\;{\rm if} 
 
\,\; 
i\in I
\end{array} } \mu(X^{\varepsilon})\\=
\sum_{\varepsilon \in \left\{ 0,1\right\}  ^{n}}\
\mu(X^{\varepsilon})\cdot \prod _{i\in 
I}\varepsilon_i=\sum_{\varepsilon \in \left\{ 0,1\right\}  ^{n}}\
\mu(X^{\varepsilon})\cdot {\bf u}^{\varepsilon }_{I} .
\end{eqnarray*}
This means that 
${\bf p} 
$ is a convex linear combination
of the vertices $ {\bf u} ^{\varepsilon}$ with weights
$\lambda_{\varepsilon}=\mu(X^{\varepsilon})$. 

Assume now that ${\bf p}$ is a convex linear combination of the 
vertices,
$${\bf p}=\sum_{\varepsilon \in \left\{ 0,1\right\}^{n}  }\lambda_{
\varepsilon}
\cdot {\bf u}^{\varepsilon }.$$
Let $\Omega =\left\{ 0,1\right\}^{n} $. 
The Kolmogorovian representation can be based on subsets 
$X_{A_{i}}=\lbrace \varepsilon \in \lbrace 0, 1\rbrace
^{n} \mid  \varepsilon_{i}=1\rbrace$. Then, for every 
$I\in S:$ 
$\bigcap_{i\in I}\ X_{A_{i}}=\lbrace \varepsilon \in 
\lbrace 0, 1\rbrace ^{n} \mid  \prod_{ i\in I 
}\varepsilon_{i}=1\rbrace$. 
Then, with the previous notation, $X^{\varepsilon}=\{\varepsilon \}$.
Let $\Sigma$ 
be the power set of 
$\Omega$. 
The measure of an arbitrary $X\in \Sigma$ is defined as  
$\mu(X)=\sum_{\varepsilon\in X}
\lambda_{\varepsilon}$. It is easy to check that this is a probability
measure. We have then
$$\mu \left( \bigcap_{i\in I}X_{A_{i}}\right)= \sum_{\varepsilon \in 
\left\{ 0,1\right\}  ^{n}}
\mu(\{\varepsilon\})\cdot {\bf u}^{\varepsilon }_{I}={\bf p}_I,$$ 
which proves that the correlation vector admits well a 
Kolmogorovian
representation.
\begin{flushright}
 $\Box$
\end{flushright}

Let us now apply the Pitowsky-Clauser-Horne inequalities 
(\ref{Eq-Clauser_Horne}) to check if
a Kolmogorovian representation exists for the `naked' quantum 
probabilities associated to
the measurement of spin directions introduced in subsection ``The Orsay 
experiments''.
$$p_{1}=tr(\hat{W} \hat{A}), p_{2}=tr(\hat{W} \hat{A'}), 
p_{3}=tr(\hat{W} \hat{B}), p_{4}=tr(\hat{W} \hat{B'}),$$ 
$$ p_{13}=tr(\hat{W} \hat{A}\hat{ B}), p_{14}=tr(\hat{W} \hat{A} 
\hat{B'}), 
p_{23}=tr(\hat{W} \hat{A'}\hat{B}), p_{24}=tr(\hat{W} 
\hat{A'}\hat{B'})$$ 
Substituting the values obtained in (\ref{qprob}) 
for the probabilities in the last inequality of 
(\ref{Eq-Clauser_Horne}) we find
$$\frac{3}{8}+ \frac{3}{8}+ \frac{3}{8}-0- \frac{1}{2}- 
\frac{1}{2}= \frac{1}{8}>0.$$ 
Consequently, 
\begin{eqnarray}
\label{notin} 
{\bf p}=\left(\frac{1}{2}, \frac{1}{2}, \frac{1}{2}, 
\frac{1}{2},\frac{3}{8}, \frac{3}{8}, 0, \frac{3}{8}\right) \not\in {\cal 
C}(n,S).
\end{eqnarray}

The {\em usual} conclusion is that the observed probabilities in the 
Orsay experiment have no Kolmogorovian representation. 
However, as Szab\'o (Szab\'o 1995a,b) pointed out, a closer analysis can 
yield
different conclusion. 

The problem, he claims, is that probabilities in (\ref{notin}) are not 
the effective relative frequencies of the observed events, the values of 
which are given in (\ref{Facts1}), but the conditional probabilities 
(\ref{qprob}). The meaning of such a conditional probability is this: the 
probability of a measurement outcome, given that the corresponding 
measurement  has been completed.  
He proved that the effective probabilities we encounter in the Orsay 
experiment, 
that is the values in (1), can be accommodated into a Kolmogorovian 
theory. For instance, he showed that the effective relative frequencies
given in (1) do no longer violate the Clauser-Horne inequalities, as
shows the following:
$$-1<\frac{3}{32}+
\frac{3}{32}+
\frac{3}{32}-0-
\frac{1}{4}- 
\frac{1}{4}= \frac{-7}{32}<0.$$  He also showed
by numerical methods ( Szab\'o 1995a,b) that the correlation vector
presented in (1), which  contains the effective probabilities of the events
$a,\,a',\, b,\, b',\,A,\,A',\,B,\,B'$ admits a Kolmogorovian 
representation. A generalisation of
Szab\'o's case (which is valid for a particular choice for the directions
$a,\,a',\, b,\, b$) to a situation which covers all possible choices of
such directions is given in ( Durt 1995 a), but suffers from a lack of
generality (it is limited to an Orsay like situation). We shall now
generalize these results, in conformity with the
Kolmogorovian censorship hypothesis mentioned in the introduction.

\section{Kolmogorovian Censorship}
\label{kc}

\subsection{Introduction, a fundamental remark.}

Let us represent by a density
operator 
$\hat{W}$ the state of the quantum system under measurement. $n$ different
measurements are carried out on it. A  measurement inspects whether the 
value of an observable 
is in a set or not. The outcome can be yes or no. 
Let $a_{i}$ denote the event 
``the $i$-th measurement has been performed". The event when
the result of
measurement $a_i$ is yes is
denoted by 
$A_{i} $. The corresponding projectors are 
$ \hat{A}_{i} $, which are given by the spectral decompositions of the 
observables. We can now formulate a remark which will appear to be
essential in the proof of the main theorem:

 {\bf Fundamental remark}: Among the
measurements there may be incompatible ones, but those that are  carried 
out simultaneously must be, and actually are, compatible.

In our 
terminology 
compatibility of measurements means that the corresponding 
operators commute. According to orthodox quantum physics only these 
measurements can be performed together (the projectors which are associated
to sets of outcomes of the same observable necessarily commute because the
basis associated to the spectral decomposition is orthocomplemented.)

The structure of the proof is the following.

We shall express in the
forthcoming subsection (4.2) the constraints imposed by the fact that the
probability distribution of the performances of the measurements is classical 
and that during simultaneous measurements, the quantum properties are represented
by  compatible
(commuting) projectors. Afterwards, we shall recall that the naked quantum
probabilities associated to compatible projectors admit a Kolmogorovian
representation (4.3). This will allow us to give a compact expression of
the effective probabilities (4.4), and to prove the main theorem (4.5).

\subsection{Constraints on the probability distribution of the performances 
of the measurements}
{\bf First assumption}:

The general correlation vector 
${\bf \tilde{p}}$ consisting of  the probabilities with which the 
measurements 
are performed, that is the one with components
$p\left(\bigwedge_{i \in I}a_i \right)$ where $I \in 
2^{\{1,...,n\}}\setminus \{\emptyset\}$ is supposed 
to be classical. This assumption is very natural, because
after all the devices used to choose which measurement is performed in the
laboratory are macroscopical devices of classically describable nature (it is also so in
the Orsay  experiments), consequently 

\begin{eqnarray}
\label{ptilde} 
{\bf \tilde {p}}=\sum_{\varepsilon \in \left\{0,1 \right\}^n} 
\kappa_{\varepsilon}{\bf u}^{\varepsilon} ,\,\,\,\,\sum_{\varepsilon 
\in 
\left\{0,1 \right\}^n} \kappa_{\varepsilon}=1, \,\,\,\, 
\kappa_{\varepsilon} \geq 0
\end{eqnarray}

{\bf Second assumption}: In conformity with the fundamental remark, we
assume that from  incompatibility of two measurements
$i,j$ (i.e. $\left[ \hat{A}_i,\hat{A}_j\right] \neq 0$) follows that they
are  not 
performed together.

This implies that a restriction on
${\bf
\tilde{p}}$ occurs. 
Let us introduce the set of ${\cal K}$ of indices which represent compatible measurements:
$${\cal K}=\left\{I\in 2^{\left\{1,...,n\right\}} \setminus \{\emptyset\} 
\mid 
\ \forall i,j \in I: \left[\hat{A}_i,\hat{A}_j\right]=0  \right\}.$$

Let $\varepsilon^I \in \{0,1\}^n$ mean the following: $\forall I \in 
2^{\left\{1,...,n\right\}} 
\setminus \{\emptyset\}:
\,\varepsilon^I _i = 1 
\Leftrightarrow i \in I$.
This is a one to one correspondence between $2^{\left\{1,...,n\right\}} 
\setminus \{\emptyset\}$ and 
$\{0,1\}^n$.

The second assumption can now be formulated as follows:
$ I \not \in {\cal K}$ $ \Rightarrow {\bf 
\tilde{p}}_I =0 .$ 
>From these assumptions follows that we can restrict the expression of
the vector ${\bf
\tilde{p}}$ as we show it now.

\begin{A}

If we assume (assumption 1) that
${\bf
\tilde{p}}$, the probability distribution of the performances of the measurements 
is classical and has the decomposition (\ref{ptilde}),
if, furthermore, we assume (assumption 2) that from  incompatibility of two
measurements
$i,j$ (i.e. $\left[ \hat{A}_i,\hat{A}_j\right] \neq 0$) follows that they
are  not 
performed together (i.e. $ I \not \in {\cal K} \Rightarrow {\bf 
\tilde{p}}_I =0 $)
then
$${\bf \tilde {p}}=\sum_{I \in {\cal K}} 
\kappa_{\varepsilon^I}{\bf u}^{\varepsilon^I} .$$
\end{A}

\paragraph{Proof:}
It is generally true and follows from decomposition  (\ref{ptilde}) that 
${\bf \tilde{p}}_I \geq \kappa_{\varepsilon^I} {\bf u}^{\varepsilon^I}_I 
= \kappa_{\varepsilon^I} $.
But, if $I \not \in {\cal K} $ then ${\bf \tilde{p}}_I =0 $ and so 
$\kappa_{\varepsilon^I}=0$.
This means exactly that 
$${\bf \tilde {p}}=\sum_{\varepsilon \in \left\{0,1 
\right\}^n} 
\kappa_{\varepsilon}{\bf u}^{\varepsilon}= \sum_{I \in {\cal K}} 
\kappa_{\varepsilon^I}{\bf u}^{\varepsilon^I} . $$

Consequently
$$ p\left(\bigwedge_{j\in I}a_{j} 
\right)=\sum_{J \in {\cal 
K}:I \subset J}\kappa_{\varepsilon^J} .$$

 \subsection{A Kolmogorovian representation for naked quantum
probabilities of compatible projectors.}

We will reproduce the proof of the well-known fact that the naked
quantum  probabilities associated 
to commuting projectors is Kolmogorovian by showing a simple explicit 
representation. 

\begin{A}
For all $J \in {\cal K}$ the correlation vector 
$${\bf \pi}^J:\  2^J \setminus \{\emptyset\} \ni I 
\rightarrow 
{\bf \pi}^J_I =tr \left(\hat{W} \prod_{i \in I} \hat{A}_i \right)$$ 
is Kolmogorovian.
\end{A}

A possible representation is the following:

$$\Omega^J= \left\{
0,1\right\}^{cardJ} $$
$$\Sigma^J=2^{\Omega^J}$$ 
$$\forall \epsilon \in \Omega^J: \mu ^J\left( \{\varepsilon \} \right)
\,=\,tr
\left(
\hat{W}\prod_{i\in J} 
\hat{A}_i^{\varepsilon 
_i}\right) $$ 
where $\hat{A}_i^0$ means the orthogonal complement of $\hat{A}_i$. 
$$X_{A_i}^J=\left\{ \varepsilon \in \Omega^J \mid \varepsilon 
_i=1\right\},\hskip 1cm i \in J $$
$$\pi_I^J = \mu \left( \bigcap_{i \in I} X_{A_i}^J\right)$$
It can be proven (Durt 1996a), by making use of the compatibility of 
the projectors
involved in the representation and of the properties of the density 
matrix that $\mu ^J$
satisfies the definition of a Kolmogorovian measure.

As an example consider the Orsay experiment.
We notice that the measurements ``$a$" 
as well as 
``$a'$" 
can be performed simultaneously with ``$b$" or ``$b'$", the 
corresponding 
operators commute. In this case, the 
correlation vectors 
\begin{eqnarray}
{\bf p}^{ab}&=&\left( tr(\hat{W}\hat{A}), tr(\hat{W}\hat{B}), 
tr(\hat{W}\hat{A}\hat{B})\right)   \nonumber \\
{\bf p}^{ab'}&=&\left( tr(\hat{W}\hat{A}), tr(\hat{W}\hat{B'}), 
tr(\hat{W}\hat{A}\hat{B'})\right)   \nonumber \\
{\bf p}^{a'b}&=&\left( tr(\hat{W}\hat{A'}), tr(\hat{W}\hat{B}), 
tr(\hat{W}\hat{A'}\hat{B})\right) \\
{\bf p}^{a'b'}&=&\left( tr(\hat{W}\hat{A'}), tr(\hat{W}\hat{B'}), 
tr(\hat{W}\hat{A'}\hat{B'})\right)   \nonumber
\end{eqnarray}
have Kolmogorovian representations: 
\begin{eqnarray}
\left(\Omega^{ab},\Sigma^{ab},\mu^{ab},\left\{X^{ab}_{A},X^{ab}_{B}\right\}\right)   
\nonumber 
\\
\left(\Omega^{ab'},\Sigma^{ab'},\mu^{ab'},\left\{X^{ab'}_{A},X^{ab'}_{B'}\right\}\right)   
\nonumber \\
\left(\Omega^{a'b},\Sigma^{a'b},\mu^{a'b},\left\{X^{a'b}_{A'},X^{a'b}_{B}\right\}\right)   
\nonumber \\
\left(\Omega^{a'b'},\Sigma^{a'b'},\mu^{a'b'},\left\{X^{a'b'}_{A'},X^{a'b'}_{B'}\right\}\right)   
\nonumber \\
\end{eqnarray}
respectively. These representations are shown in Table 1, 
corresponding to the particular
choice of directions a, a', b' b', made at the beginning of this work.

\begin{table}[h]
\begin{center} $a \cap b$ \hskip 6.2 truecm $a \cap b'$\end{center}
\begin{center}
\begin{tabular}{|c|c|}\hline
$A \cap B$ & $A \cap \neg B$ \\
$\qquad {3\over 8}\qquad$ & $\qquad{1 \over 8}\qquad$ \\
\hline 
$\neg A \cap B$ & $\neg A \cap \neg B$ \\
${1 \over 8}$ & ${3 \over 8}$ \\ \hline
\end{tabular} $\quad$
\begin{tabular}{|c|c|}\hline
$A \cap B'$ & $A \cap \neg B'$ \\
$\qquad {3\over 8}\qquad$ & $\qquad{1 \over 8}\qquad$ \\
\hline 
$\neg A \cap B'$ & $\neg A \cap \neg B'$ \\
${1 \over 8}$ & ${3 \over 8}$ \\ \hline
\end{tabular}
\end{center}
\begin{center} $a' \cap b$ \hskip 6.2 truecm $a' \cap b'$ \end{center} 
\begin{center}
\begin{tabular}{|c|c|}\hline
$A' \cap B$ & $A' \cap \neg B$ \\
$\qquad 0\qquad$ & $\qquad{1\over 2}\qquad$ \\
\hline 
$\neg A' \cap B$ & $\neg A' \cap \neg B$ \\
${1\over 2}$ & $0$ \\ \hline
\end{tabular}
$\quad$
\begin{tabular}{|c|c|}\hline
$A' \cap B'$ & $A' \cap \neg B'$ \\
$\qquad {3 \over 8}\qquad$ & $\qquad{1 \over 8}\qquad$ \\
\hline 
$\neg A' \cap B'$ & $\neg A' \cap \neg B'$ \\
${1 \over 8}$ & ${3\over8}$ \\ \hline
\end{tabular}
\end{center}
\caption{The Kolmogorovian representations of the ``naked'' Orsay
frequencies for  the four choices
of experimental arrangements.}
\end{table}

\subsection{A compact expression of the effective probabilities.}
As we already emphasised, the ``naked'' quantum 
probabilities $tr \left( \hat{W} \hat{A}_{1}\right) $,  $tr \left( 
\hat{W} 
\hat{A}_{2}\right) $,
..., $tr \left( \hat{W} \hat{A}_{i} \hat{A}_{j} \right) $ are conditional 
probabilities. In order to get the probabilities of the outcomes we must 
multiply these values by the 
probabilities of the performance of the corresponding measurements.
That is, the effective probability of, for instance $ A_{1} $, is 
$$p(A_{1})=p\left(A_{1} \wedge a_{1} \right) = p\left(A_{1} \mid a_{1} 
\right)\cdot p(a_{1}) =p(a_{1})\cdot tr\left( \hat{W} 
\hat{A_{1}}\right). $$
You may disagree with the usage of the classical form of conditional 
probability, but whenever the 
``naked'' quantum probabilities are testified in an experiment they are 
compared with the relative 
frequences of the outcomes {\em relative to} the performance of the 
measurement.
Similarily, for the conjunctions of outcomes we have
$$p\left(\bigwedge_{i\in I}A_{i}\right) = p\left(\bigwedge_{i\in I}a_{i} 
\right) \cdot tr \left( \hat{W} \prod_{i\in I}\hat{A}_{i}\right) .$$ 
For any $I_{1},I_{2}\subset 2^{\left\{1,...,n\right\}}\setminus 
\emptyset$:
\begin{eqnarray}
\label{metszetval} 
p\left(\left(\bigwedge_{i\in I_1}A_{i}\right)\wedge 
\left(\bigwedge_{j\in 
I_{2}}a_{j} \right)\right)= p\left(\left(\bigwedge_{i\in I_1}A_{i} 
\right)\wedge \left(\bigwedge_{j\in I_1 \cup I_{2}}a_{j} 
\right)\right) \nonumber\\=p\left(\bigwedge_{i\in I_1}A_{i} \mid 
\bigwedge_{j\in 
I_1 \cup 
I_{2}}a_{j} \right)\cdot p\left(\bigwedge_{j\in I_1 \cup I_{2}}a_{j} 
\right) \nonumber\\= 
p\left(\bigwedge_{j\in I_1 \cup I_{2}}a_{j} \right) \cdot tr \left( 
\hat{W} 
\prod_{i\in I_1}\hat{A}_{i}\right) \nonumber.
\end{eqnarray}
We used, for the first equality, the fact that it is impossible to 
observe the event
$A_i$ without performing the experiment $a_i$. Notice, that these 
expressions are 
valid even if the measurements are not compatible, then both sides are 
zero.

Thus we are ready to prove the main theorem.

\subsection{Proof of the main theorem.}

We claim that the generalized
correlation vector
${\bf p}$ that contains  the probabilities of  events 
$A_{1} $,  $ A_{2} $, ..., $ A_{n} $, $a_{1} $, $ a_{2} $, ..., $ a_{n} $ 
and all their conjunctions is 
Kolmogorovian.  

\begin{T}
If  ${\bf \tilde{p}}$, the probability distribution of the
performances of the measurements is
classical, has the 
decomposition (\ref{ptilde}), and if from the incompatibility of two 
measurements $i,j$
(i.e. $\left[ \hat{A}_i,\hat{A}_j\right] \neq 0$) follows that they are 
not 
 performed 
together then 
the effective probabilities associated to the events $A_{1} $,  $ A_{2} 
$, ..., $ A_{n} $,
$a_{1}
$,
$ a_{2}
$, ...,
$ a_{n} $  and all their conjunctions admit a Kolmogorovian 
representation.
\end{T}

\paragraph{Proof:}

In order to prove the theorem, we shall build an explicit Kolmogorovian 
representation for
the effective correlation vector.  Let us introduce the disjoint union of 
$\Omega^J$-s defined above:
$$\Omega=\bigcup_{J\in {\cal K}}\Omega^J,$$
and let $\Sigma$ be the $\sigma$-algebra on $\Omega$ generated by 
$\bigcup_{J\in {\cal K}}\Sigma^J$. We can extend the measures $\mu^J$ 
onto $\Sigma$ in a 
natural way:
$$\mu^J(X)=\mu^J\left(X\cap \Omega^J\right),\,\,\,\,\,(X \in \Sigma)$$
and define a new probability measure as 
$$\mu=\sum_{J \in {\cal K}} \kappa_{\varepsilon ^J }\cdot \mu^J.$$ It is 
easy to check that
this is a Kolmogorovian measure. The representative sets of $A_i$-s 
and $a_j$-s are
constructed as 
$$X_{A_i}=\bigcup_{J\in {\cal K}:i\in J} 
X_{A_i}^J,\,\,\,\,\,X_{a_j}=\bigcup_{J\in 
{\cal K}:j \in J} \Omega^J$$
According to the definition, $X_{A_i}\subset X_{a_i}$ for every $i\leq 
n$, 
and if for some $i$ and $j$ the respective measurements are not 
compatible then $X_{A_i}\cap X_{a_j}=\emptyset$ because $i$ and $j$ 
cannot be in the same $J$ of ${\cal K}$. So $\mu$ gives $0$ probability 
for them. 
For arbitrary $I_1,I_2\in2^{\{1,...,n\}}\setminus 
\left\{\emptyset \right\}$:
\begin{eqnarray*}
\bigcap_{i\in I_1} X_{A_i}= \bigcap_{i\in I_1} \bigcup_{J_1\in {\cal 
K}:i\in J_1} 
X^{J_1}_{A_i} =\\ 
\left\{ x \mid (\forall i \in I_1)(\exists J_1 \in {\cal K}):\, i \in J_1 
\,\, and \,\,x \in X^{J_1}_{A_i} 
\right\}=\\
\left\{ x \mid \exists J_1 \in {\cal K}:\, I_1 \subset J_1 \,\, and \,\,x 
\in \bigcap_{i \in 
I_1}X^{J_1}_{A_i} \right\}=\\
\bigcup_{J_1\in {\cal K}:I_1 \subset 
J_1} 
\left(\bigcap_{i\in I_1} X^{J_1}_{A_i}\right)
\end{eqnarray*}
and
\begin{eqnarray*}
\bigcap_{j\in I_2} X_{a_j}=\bigcap_{j\in I_2} \bigcup_{J_2\in {\cal 
K}:j\in J_2} \Omega^{J_2}=\\
\left\{ x \mid (\forall j \in I_2)(\exists J_2 \in {\cal K}):\, j \in J_2 
\,\, and \,\,x \in \Omega^{J_2} 
\right\}=\\
\left\{ x \mid \exists J_2 \in {\cal K}:\, I_2 \subset J_2 \,\, and \,x 
\in \Omega^{J_2} \right\}=\\
\bigcup_{J_2\in {\cal K}:I_2 \subset 
J_2} \Omega^{J_2} 
\end{eqnarray*}
This means that
\begin{eqnarray}
\label{metszet}
\left(\bigcap_{i\in I_1} X_{A_i}\right) \cap \left(\bigcap_{j\in I_2} 
X_{a_j} \right)= \bigcup_{J\in {\cal K}:I_1 \cup I_2 \subset 
J} 
\left(\bigcap_{i\in I_1} X^{J}_{A_i} \right).
\end{eqnarray}
If $I_1\cup I_2$ contains incompatible 
measurements then (\ref{metszet}) gives $\emptyset$, because there is no 
$ J\in {\cal K}$
containing $I_1\cup I_2$. But if $ I_1\cup I_2 \in {\cal K}$ then
\begin{eqnarray*}
\mu\left(\left(\bigcap_{i\in I_1} X_{A_i}\right) \cap 
\left(\bigcap_{j\in I_2} X_{a_j}\right) \right)=\\
  \mu\left(\bigcup_{J\in {\cal K}:I_1 \cup I_2 \subset 
J} 
\left(\bigcap_{i\in I_1} X^{J}_{A_i} \right) \right)=\\  \sum_{J \in 
{\cal 
K}:I_1 \cup I_2 \subset J}\kappa_{\varepsilon^J}  \cdot \mu^J 
\left(\bigcap_{i\in I_1} X_{A_i}^J \right) =\\  \left( \sum_{J \in {\cal 
K}:I_1 \cup I_2 \subset J}\kappa_{\varepsilon^J} \right) \cdot 
tr\left(\hat{W} 
\prod_{i\in I_1} \hat{A}_i \right)=
\\  p\left(\bigwedge_{j\in I_1 \cup I_{2}}a_{j} 
\right) \cdot tr \left( \hat{W} \prod_{i\in I_1}\hat{A}_{i}\right).
\end{eqnarray*}
which is exactly the same as the effective probability of the event 
$\left(\bigwedge_{i\in
I_1}A_{i}\right)\wedge 
\left(\bigwedge_{j\in 
I_{2}}a_{j} \right)$ obtained at the beginning of this section.

We get similarily that for some $I \in {\cal K}$:
$$\mu \left( \bigcap_{i\in I}X_{a_i} \right)=p\left( \bigwedge_{i\in 
I}a_i 
\right)$$
and
$$\mu \left( \bigcap_{i\in I}X_{A_i} \right) =p\left( 
\bigwedge_{i\in 
I}a_i 
\right) \cdot 
tr \left( \hat{W} \prod_{i\in I} \hat{A}_{i} \right),$$ showing the 
theorem.
\begin{flushright}
 $\Box$
\end{flushright}

This last form shows that if we denote the correlation vector $I \mapsto 
tr \left( \hat{W} \prod_{i\in I}\hat{A}_{i}\right)$ with the symbol 
${\bf \pi}$, 
then we have a Kolmogorovian representation for the vector $
{\bf \tilde{p}} \cdot {\bf \pi}$, in accordance with the notations of the 
Kolmogorovian
censorship hypothesis.

For example, the Kolmogorovian representation associated to the Bell-like 
experiment
described at the beginning of this work is given in Table 2.
\begin{table}[bht]
\begin{center}
\begin{tabular}{|c|c|c|c|}\hline
$A \cap B$ & $A \cap \neg B$ & $A \cap B'$ & $A \cap \neg B'$ \\
$\qquad {3 \over 32}\qquad$ & $\qquad{1 \over 32}\qquad$ &
$\qquad{3 \over 32}\qquad$ & $\qquad{1 \over32}\qquad$ \\
\hline  $\neg A \cap B$ & $\neg A \cap \neg B$ & $\neg A \cap B'$ & $\neg 
A \cap \neg
B'$ \\ 
${1 \over 32}$ & ${3 \over 32}$ & $\qquad{1 \over 32}\qquad$ & $\qquad{3 
\over 32}\qquad$
\\ \hline 
$A' \cap B$ & $A' \cap \neg B$ & $A' \cap B'$ & $A' \cap \neg B$ \\
$\qquad 0\qquad$ & $\qquad{1\over 8}\qquad$ &
$\qquad{3 \over 32}\qquad$ & $\qquad{1 \over 32}\qquad$ \\
\hline  $\neg A' \cap B$ & $\neg A' \cap \neg B$ & $\neg A' \cap B'$ & 
$\neg A' \cap
\neg B'$ \\ 
${1 \over 8}$ & $0$ & $\qquad{1 \over 32}\qquad$ & $\qquad{3 \over 
32}\qquad$ \\
\hline 
\end{tabular}
\end{center}
\caption{The Kolmogorovian representation of the Orsay effective 
frequencies.}
\end{table}

\section{Conclusion.}
We shall not, in our conclusion, discuss in details the possible 
interpretations of our result.
Some of them can be found in other articles.
We just mention that this representation can be used for the proof of
the existence of a local  deterministic hidden variable model (Durt 1995,
1996a,b, Szab\'o 1995 a,b).

The aim of the article was to prove Szab\'o's Kolmogorovian Censorship 
hypothese. We were able to 
do this by taking into account only observable events, that is, the 
performances of the measurements 
and the beeps of the detectors. We supposed that the incompatible 
measurements are not carried out 
together and that the probability distribution of the   
performances of the measurements is classical. This way 
the quantum probabilities appear as classical conditional probabilities. 
It is useful to remark that 
because the events of carrying out the measurements are taken into the 
event algebra, in hidden 
variable models the choice of an experiment (the 
choice of the
direction of the magnet in our example) is dependent on the value taken 
by the hidden
variable. This has important consequences for the question of determinism 
and free-will (Durt 1995, 1996a,b;  Szab\'o 1995a,b).

\section{Acknowledgements.}
We wish to thank L. E. Szab\'o for the numerous fruitful discussions held 
in Brussels and 
Budapest with everyone of us. One of us (T. Durt) was supported by the 
Federale
Diensten voor Wet., Techn. en Cult. Aang. in the
framework of IUAP-III n$^{\rm o}9$ and by the Interuniversitaire 
Instituut voor Kern
Wetenschappen during the realisation of this work.

\section*{References}

Accardi, L., and Fedullo, A., (1982): On the statistical
meaning of the complex numbers in quantum mechanics, {\it Nuovo Cimento\/}
{\bf 34}, 161.

Aspect, A., Dalibard, P., and Roger, G., (1981): Experimental tests of
realistic local theories via Bell's theorem, {\it  Phys. Rev. Lett.\/} 
{\bf 47}, 
460.

Bell, J.S., (1964): On the EPR paradox, {\it  Physics \/}{\bf 1}, 195.

Clauser, J.F., and Horne, M.A., (1974): Experimental consequences of
objective local theories, {\it  Phys. Rev.\/}  {\bf D10}, 526.

Durt, T., (1995): Three interpretations of the violation of Bell's
inequalities, to be published in the
Foundations of Physics (accepted for publication in November
1996).

Durt, T., (1996a): {\it  From quantum to classical, a toy model.\/}, 
Doctoral
thesis, January 1996. 

Durt, T., (1996b): Why God might play dice, to be published in the 
{\it  Int. J. Theor. Phys.\/}.

Pitowsky, I., (1989): {\it  Quantum probability. Quantum logic\/}, 
Lecture Notes in Physics {\bf 321}, Springer Verlag, Berlin

Szab\'o, L  E.., (1993): On the real meaning of Bell's theorem,  { \em  
Foundations of Physics Letters}, { \bf 6}, 191.

Szab\'o, L. E., (1995a):  Quantum Mechanics in an Entirely Deterministic 
 
Universe,  { \em Int. J. Theor. Phys.} { \bf 34}, 1751-1766.

Szab\'o, L. E., (1995b): Is quantum mechanics compatible with a 
deterministic  
universe? Two interpretations of quantum probabilities,  { \em  
Foundations of Physics Letters},  { \bf 8}, 421-440.

\end{document}